\documentclass[conference]{IEEEtran}

\usepackage{xurl}
 \usepackage{xcolor} 
\usepackage{bm}
\usepackage{amsmath}
\usepackage{lipsum}
\usepackage{graphicx}
\ifCLASSOPTIONcompsoc
    \usepackage[caption=false, font=normalsize, labelfont=sf, textfont=sf]{subfig}
\else
\usepackage[caption=false, font=footnotesize]{subfig}
\fi

\usepackage[ruled,linesnumbered]{algorithm2e}
\SetKwInput{KwInput}{Input}      
\SetKwInput{KwOutput}{Output}    
\SetKwInput{KwInit}{Initialization}
\SetKwInput{KwPara}{Parameters} 
\SetKwInput{KwReturn}{return}

\usepackage [english]{babel}
\usepackage [autostyle, english = american]{csquotes}
\MakeOuterQuote{"}
\IEEEoverridecommandlockouts

\ifCLASSINFOpdf
\else
\fi
\begin{document}

\title{Optimal UAV Hitching on Ground Vehicles}

\author{Lihua Ruan, Lingjie Duan, and Jianwei Huang$^\star$ \thanks{This work is supported by the Shenzhen Institute of Artificial Intelligence and Robotics for Society, and the Presidential Fund from the Chinese University of Hong Kong, Shenzhen.

Lihua Ruan is with School of Science and Engineering, Shenzhen Institute of Artificial Intelligence and Robotics for Society, The Chinese University of Hong Kong, Shenzhen, and University of Science and Technology of China, Hefei, China (email:ruanlihua@cuhk.edu.cn).

Lingjie Duan is with the Engineering Systems and Design Pillar, Singapore University of Technology and Design, Singapore, and the Shenzhen Institute of Artificial Intelligence and Robotics for Society, Shenzhen, China (email:lingjie\_duan@sutd.edu.sg).

Jianwei Huang is with School of Science and Engineering, Shenzhen Institute of Artificial Intelligence and Robotics for Society, The Chinese University of Hong Kong, Shenzhen, Shenzhen, 518172, China ($^\star$corresponding author, e-mail: jianweihuang@cuhk.edu.cn).
}
}

\maketitle

\begin{abstract} Due to its mobility and agility, unmanned aerial vehicle (UAV) has emerged as a promising technology for various tasks, such as sensing, inspection and delivery. However, a typical UAV has limited energy storage and cannot fly a long distance without being recharged. This motivates several existing proposals to use trucks and other ground vehicles to offer riding to help UAVs save energy and expand the operation radius. We present the first theoretical study regarding how UAVs should optimally hitch on ground vehicles, considering vehicles' different travelling patterns and supporting capabilities. For a single UAV, we derive closed-form optimal vehicle selection and hitching strategy. When vehicles only support hitching, a UAV would prefer the vehicle that can carry it closest to its final destination. When vehicles can offer hitching plus charging, the UAV may hitch on a vehicle that carries it farther away from its destination and hitch a longer distance. The UAV may also prefer to hitch on a slower vehicle for the benefit of battery recharging. For multiple UAVs in need of hitching, we develop the max-saving algorithm (MSA) to optimally match UAV-vehicle collaboration. We prove that the MSA globally optimizes the total hitching benefits for the UAVs.

\end{abstract}


%
\IEEEpeerreviewmaketitle

\section{Introduction}
\subsection{Motivations}
Unmanned aerial vehicles (UAVs), colloquially known as drones, reinvent many industrial solutions in recent years \cite{DroneSurvey}. For example, in logistics, Amazon deploys Prime Air, a drone-based fast delivery service in directing parcels to home. In communications, AT$\&$T uses "Flying COWs" drones to offer LTE services. In the energy industry, drone is now a major tool for facility inspection. When using UAVs for these various tasks, a major challenge lies in UAV's limited energy supply.

Common commercial UAVs, e.g., DJI and Parrot professional drones, typically can sustain a flight up to around 30 mins. Adding loads to UAVs further shortens the flight time. The limited energy supply prevents UAVs from flying over wide areas or persisting operation in the sky. This energy problem is unavoidable even with optimized UAV trajectory and energy consumption. One way to support UAV task execution over a wide range is to exploit ground vehicles to carry UAVs. For example, logistics firms use their delivery trucks as UAV carriers \cite{AmazonPA}. The authors in
\cite{UAVBus1} proposed to leverage public transportation, i.e., buses and trains, to carry UAVs for city-wide surveillance. Amazon Prime Air reveals their initiative to forge drone-truck riding agreements with shipping companies so that their drones can hitch on its collaborators' trucks \cite{AmazonRide}. Drones can identify trucks via special markings and then inform truck drivers for hitching \cite{AmazonPatent}. 

Given the important role of vehicles in assisting UAVs to save energy and sustain operation, we solve \textit{how UAVs should choose among various ground vehicles when needing hitching} and \textit{how far to hitch on the vehicle} to reduce their consumption. Note that the autonomous UAV landing on static or moving vehicles is already technically mature \cite{UAVLanding}. The key requirement in landing a UAV on vehicle is the landing marker for the purpose of positioning. In addition to hitching, we further consider the case where vehicles can offer energy charging to UAVs. Existing vehicles can already charge drones with the direct current ouput \cite{DJICarCharger}. On-vehicle charging pad and dock are also available \cite{Wibotic}. Hence, in this study, we consider both hitching-only and hitching-plus-charging vehicles.

\subsection{Contributions}

To our best knowledge, this work is the first to address how UAVs should optimally hitch on ground vehicles for UAVs' energy saving. Our key novelty and contributions are listed:
\begin{itemize}
\item \emph{Novel study of UAV-vehicle collaboration}: We present the first theoretical study regarding how UAVs should optimally choose their vehicle collaborators to hitch and how far to hitch, considering vehicles' heterogeneous travel directions, speeds and capabilities in charging UAVs.

\item \emph{Optimal UAV hitching on hitching-only vehicles}: We derive closed-form solutions of UAV's choice of vehicle and hitching distance. Without charging, the UAV will only hitch on a vehicle that brings it closer to its destination.
\item \emph{Optimal UAV hitching on hitching-plus-charging vehicles}: We derive the UAV's closed-form optimal hitching choice on hitching-plus-charging vehicles. With charging, the UAV may hitch on a vehicle taking it further away from its destination. The UAV may also prefer a slower vehicle for the benefit of battery charging. 

\item \emph{Optimal matching UAVs with ground vehicles}: For multiple UAVs needing hitching, we develop the max-saving algorithm (MSA) to optimally match UAV-vehicle pairs as well as optimize each UAV's hitching distance. The MSA maximizes the energy saving for all UAVs.

\end{itemize}

\subsection{Related Work}
The state-of-the-art literature related to using vehicles to support UAVs mostly focused on the aspect of trajectory planning. Several studies such as in \cite{VRPD1} and \cite{VRPD2} investigated the Vehicle Routing Problem with Drones (VRPD) for jointly optimizing the trajectories of drone and truck in delivery. The authors in \cite{VRPD3} studied a variant of the VRPD, which is to plan the trajectory of a recharging vehicle to accompany UAV in information collection. References \cite{UAVBus1} focused on optimizing the UAV's trajectory 
considering bus timetables. Our work investigates the new form of UAV-vehicle collaboration in hitching. Though a recent work \cite{Hitching} also studies the UAV hitching, it focuses on online algorithm design and limits to a single path scenario instead of the UAV-vehicle collaboration. In this paper, along with the optimal UAV hitching solutions, we capture the critical vehicle features to be considered when UAVs seek hitchhiking, including speed, travel direction and the capability to charge UAVs, and characterize the impacts of these features on the UAV-vehicle collaboration.

In the remainder of this paper, Section II presents our system model. The optimal UAV hitching solutions are in Section III. Section IV presents the MSA. Results are in Section V. We summarize in Section VI.

\section{System Model}

\begin{figure}
    \centering
    \hspace*{-0.4cm}\includegraphics[width=1.1\linewidth]{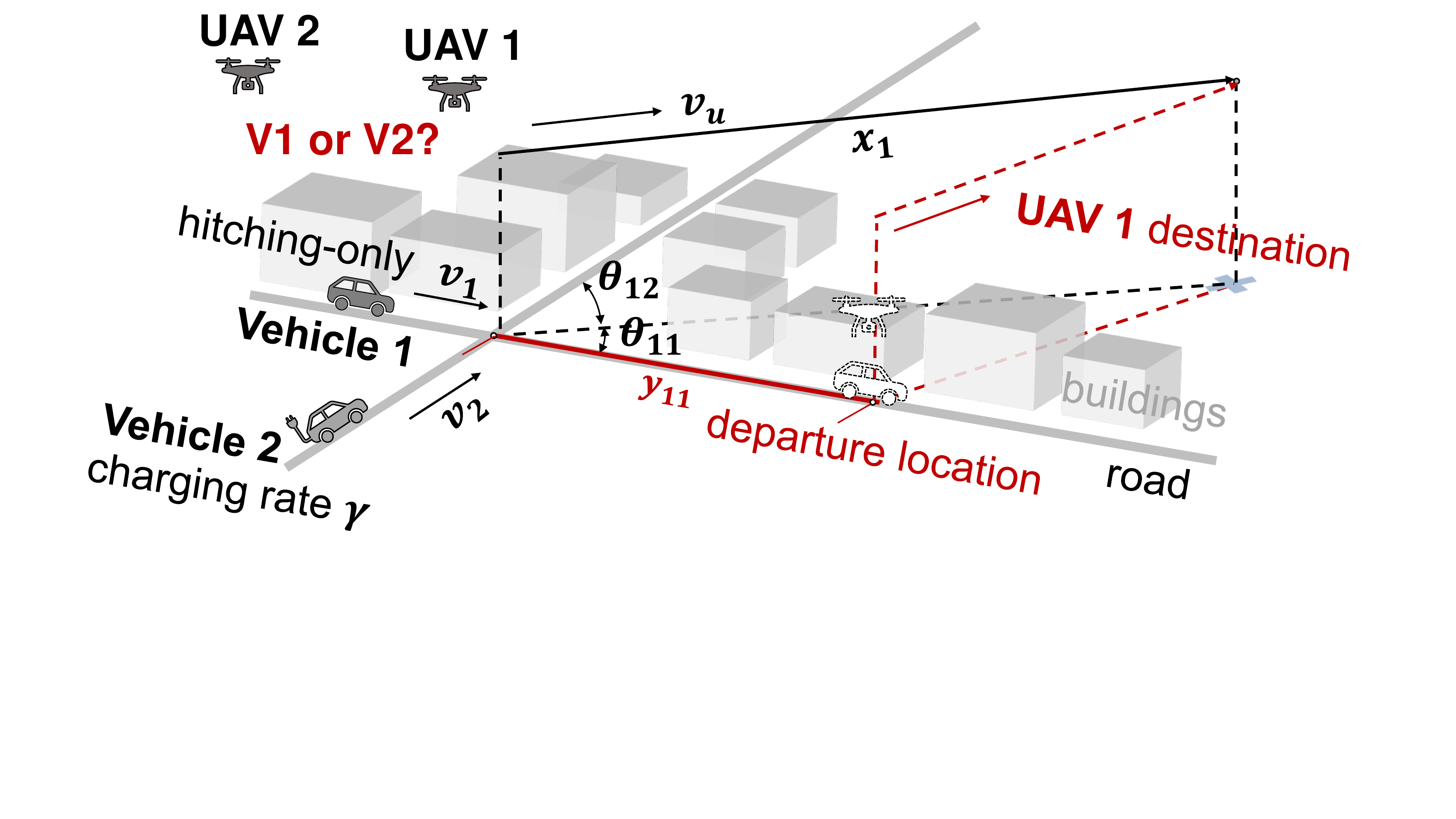}
    \vspace{-70pt}
    \caption{Model of UAV-vehicle hitching collaboration. Each UAV can either fly to its destination or hitch on an available vehicle nearby for part of the trip. The questions are which vehicle to choose and how far to hitch.}
    \label{fig:system model}
\end{figure}

Fig. \ref{fig:system model} illustrates the UAV-vehicle hitching collaboration scenario. We consider a set $\mathcal{I}=\{1, \ldots, I\}$ of UAVs that need hitching and $\mathcal{J} = \{1, \ldots, J\}$ of vehicles that are available to support. The UAVs and vehicles are at the same location such as shown in Fig. \ref{fig:system model}. A UAV $i \in \mathcal{I}$ intends to travel to its destination that is $x_i$ distance away from its current location within time $D_i$. The flying speed of UAV $i$ is denoted by $u_i$. The UAV $i$ can reach its destination by the following means:
\begin{itemize}
    \item \textbf{Choice 1:} UAV $i$ can directly fly over distance $x_i$ to its destination. This serves as the energy consumption benchmark. 
    \item \textbf{Choice 2:} UAV $i$ can hitch on one of the ground vehicles, e.g., vehicle $j \in \mathcal{J}$, for some distance before flying to its destination. We represent the hitching distance as $y_{ij}$.
\end{itemize}

Ground vehicles generally travel in different directions and speeds. The travel direction of a vehicle can be different from that of a UAV. We denote the deviation of vehicle $j$'s travel direction from UAV $i$'s destination direction as $\theta_{ij}$, $0\leq\theta_{ij}\leq\pi$. The speed of vehicle $j$ is denoted by $v_j$. Importantly, we consider some vehicles can offer charging to UAVs, while the others only support hitching. We use $\gamma_j$ to represent the UAV charging rate offered by the vehicle $j$, where $\gamma_j = 0$ indicates hitching-only and $\gamma_j > 0$ indicates hitching-plus-charging service.

In this paper, we try to answer the following questions:
\begin{itemize}
    \item \textbf{Question 1:}  Given a single UAV $i $ and vehicle $j$, should the UAV hitch on the vehicle? If yes, how far to hitch?
    \item \textbf{Question 2:}  Given multiple vehicles to choose from, which vehicle should the UAV $i$ choose to hitch on?
    \item \textbf{Question 3:} Given multiple UAVs and multiple vehicles in $\mathcal{J}$, how should the UAVs optimally hitch on vehicles to maximize the overall benefit to the entire UAV group? 
\end{itemize}

Take UAV 1 in  Fig. \ref{fig:system model} as an example. There is a hitching-only Vehicle 1 (V1) that can carry it close to its destination in direction $\theta_{11}$ and there is a hitching-plus-charging Vehicle 2 (V2) in direction $\theta_{12} > \theta_{11}$ that offers charging. Should the UAV 1 hitch on V1 to get close to its destination or on V2 for charging? And for how long on the chosen vehicle in order to maximize the energy reduction while meeting the delay constraint? What if both UAV 1 and UAV 2 need hitching and need to choose among V1 and V2?
\setlength{\textfloatsep}{10pt}

\subsection{UAV's travel time and energy consumption}

We first model the travel time and energy consumption when UAV $i$ with parameters $(x_i, u_i, D_i)$ hitches on vehicle $j$ with parameters $(\theta_{ij}, v_j,\gamma_j)$ for a distance of $y_{ij}$. We consider the UAV's additional flying distance due to landing on and taking off from a vehicle is negligible compared to the hitching distance. As reported in \cite{UAVLanding} and \cite{UAVLanding2}, the UAV can complete landing on a fast-moving vehicle in a few tens of meters, and within a few meters if the vehicle is moving slowly.

With hitching distance $y_{ij}$, we compute the total traveling time from the UAV's starting position to its destination as: 
\begin{equation}
\displaystyle 	T_{ij}(y_{ij}) =  \frac{y_{ij}}{v_j} + \frac{\sqrt{x_i^2-2x_iy_{ij}\cos\theta_{ij}+y_{ij}^2}}{u_i},
\end{equation}
where $T_{ij}(y_{ij})$ includes both the hitching time $y_{ij}/v_j$ on the vehicle, and the flying time over a distance of $\sqrt{x_i^2-2x_iy_{ij}\cos\theta_{ij}+y_{ij}^2}$ to the the destination.

The UAV saves energy during hitching and consumes energy in flying once it departs from the vehicle. If the vehicle offers charging, the UAV can be compensated with energy during hitching. Hence, we denote the energy consumption of UAV $i$ by $E_{ij}(y_{ij})$, expressed as\footnote{For the ease of analysis, we assume that the UAV battery capacity is large enough so that it can charge as much as possible on the vehicle. Due to the space limitation, we provide additional analysis considering the UAV's energy charging state in Appendix A.}:
\begin{equation}
E_{ij}(y_{ij}) = 
\displaystyle  \frac{\sqrt{x_i^2-2x_iy_{ij}\cos\theta_{ij}+y_{ij}^2}}{u_i}-\frac{\gamma_j}{v_j}y_{ij}.
\end{equation}
On the right hand side of (2), the first term is the energy consumption of the UAV flying to its destination from the starting position. The second term is the amount of energy charged to the UAV with a charging rate $\gamma_j$.

Typically, the UAV needs to balance its travel time and energy consumption \cite{Tradeoff1}\cite{Tradeoff2}. Even after considering the time deadline $D_i$, the UAV may still prefer to reach its destination faster. For example, Amazon promises a 30-min deadline of drone delivery but it still endeavors to reach customers as soon as possible \cite{AmazonPA}. Another example is in emergency situations, where we expect the UAV to arrive an emergency location as soon as possible. Thus, we assume that the UAV trades off $T_{ij}(y_{ij})$ in (1) and $E_{ij}(y_{ij})$ in the form of $C_{ij}(\omega,y_{ij})$:
\begin{equation}
\displaystyle 	C_{ij}(\omega,y_{ij}) =\omega E_{ij}(y_{ij}) + (1-\omega)T_{ij}(y_{ij}),
\end{equation}
where $\omega \in [0,1]$ is the weight on energy over travel time. A larger $\omega$ maps a higher concern on its energy. $C_{ij}(\omega,y_{ij})$ is the consumption of the UAV $i$ by hitching on vehicle $j$.

In the benchmark case where the UAV $i$ flies to its destination, we can derive the consumption by substituting $y_{ij} = 0$ in (3). The consumption is irrelevant to $j$  as follows:
\begin{equation}
    C_{i0} = x_i/u_i.
\end{equation}

\subsection{Optimization problem}

For UAVs in $\mathcal{I}$ and vehicles in $\mathcal{J}$, we formulate the optimal UAV hitching problem as follows:
\begin{subequations}
\begin{alignat}{3}
\text{P1:} \,\, & \underset{\bm{b},\bm{y}}{\text{maximize}}   &\quad& \sum_{i=1}^{I}\sum_{j=1}^{J}(C_{i0} - C_{ij}(\omega,y_{ij}))b_{ij} \\
& \text{subject to}   &\quad&  \sum_{i=1}^{I} b_{ij} \leq 1, \forall j; \sum_{i=1}^{J} b_{ij} \leq 1, \forall i,\\
&  &\quad&  b_{ij} \in \{0,1\}, \,\,\,\,\,\quad\quad \forall i,j,\\
&  &\quad&  y_{ij}(1-b_{ij}) = 0, \quad \forall i,j,\\
&  &\quad&  y_{ij} \geq 0, \quad\quad \forall i,j,\\
&  &\quad&  T_{ij}(y_{ij})\leq D_i, \quad \forall i,j,
\end{alignat}
\end{subequations}
where $C_{i0} - C_{ij}(\omega,y_{ij})$ is the consumption reduced via hitching. Clearly, the objective in (5a) is to maximize the benefits of hitching collaboration among $I$ UAVs and $J$ vehicles. In specific, $b_{ij}$ in (5a) is the binary variable, indicating whether UAV $i$ hitches on vehicle $j$ ($b_{ij} = 1$) or not  ($b_{ij} = 0$). Constraint (5b) ensures that each UAV selects at most one vehicle and each vehicle accommodates at most one UAV \footnote{This can be extended to the case where vehicles can carry different number of UAVs. We present the details in Appendix C.}. 
Further, constraint (5d) ensures that $y_{ij}$ can only be positive when the hitching collaboration happens. The last constraint (5f) ensures the UAV $i$ can reach its destination within $D_i$.

In Problem P1, multiple UAVs need to jointly optimize their hitching choices. To solve Problem P1, we apply two-stage optimization: 
\begin{itemize}
    \item \textbf{Stage 1:} We first focus on a single UAV $i$'s hitching choice for its own benefit, regardless of its impact to other UAVs. We optimize $y_{ij}$ between a UAV $i$ and a vehicle $j$. This allows us to compare the UAV $i$'s hitching benefits when it faces multiple vehicles. This provides the answers to Questions 1 and 2 raised earlier.
    \item \textbf{Stage 2:} After characterizing the hitching benefit of between each UAV $i$ and vehicle $j$ solved in Stage 1, we optimize $\{b_{ij}\}$ among $I$ UAVs and $J$ vehicles considering a bipartite graph matching. This optimally solves problem P1 and answers Question 3 raised earlier. 
\end{itemize}

\section{Stage I: Optimal UAV Hitching Strategy}
\begin{figure*}[!t]
\minipage{0.32\textwidth}
\centering
\includegraphics[width=0.9\linewidth, height = 1.68in]{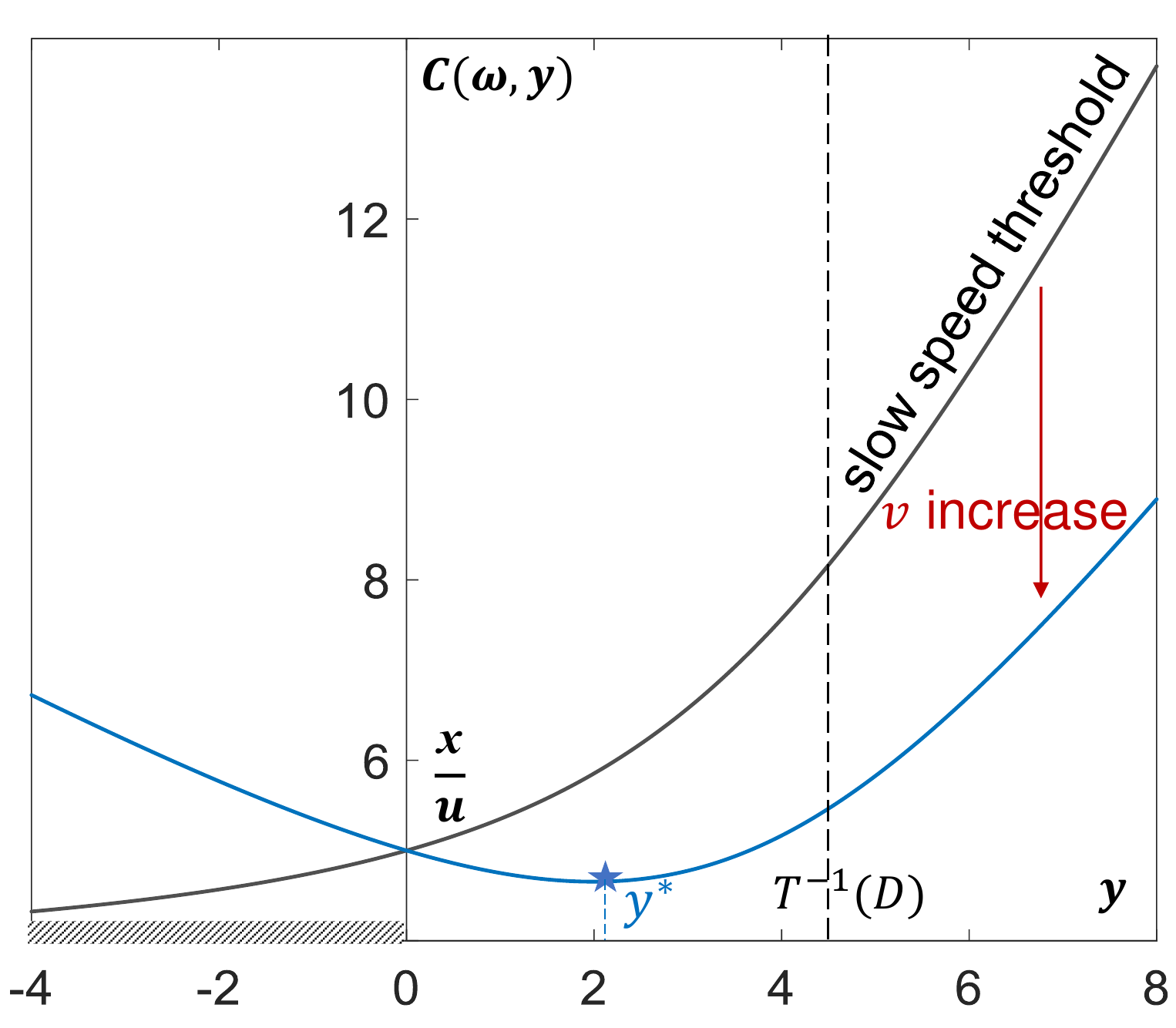}%
\caption{Vehicle speed $v$ and $C(\omega,y)$. A hitching-only vehicle with a speed lower than the slow speed threshold is not eligible. The parameters are $x = 5 \, km, \omega = 0.8$, and $u = 60 \, km/h$ with the slow speed threshold $v = 12 \, km/h$.}
\label{fig:speed}
\endminipage\hfil
\minipage{0.32\textwidth}
\centering
\includegraphics[width=0.86\linewidth, height = 1.68in]{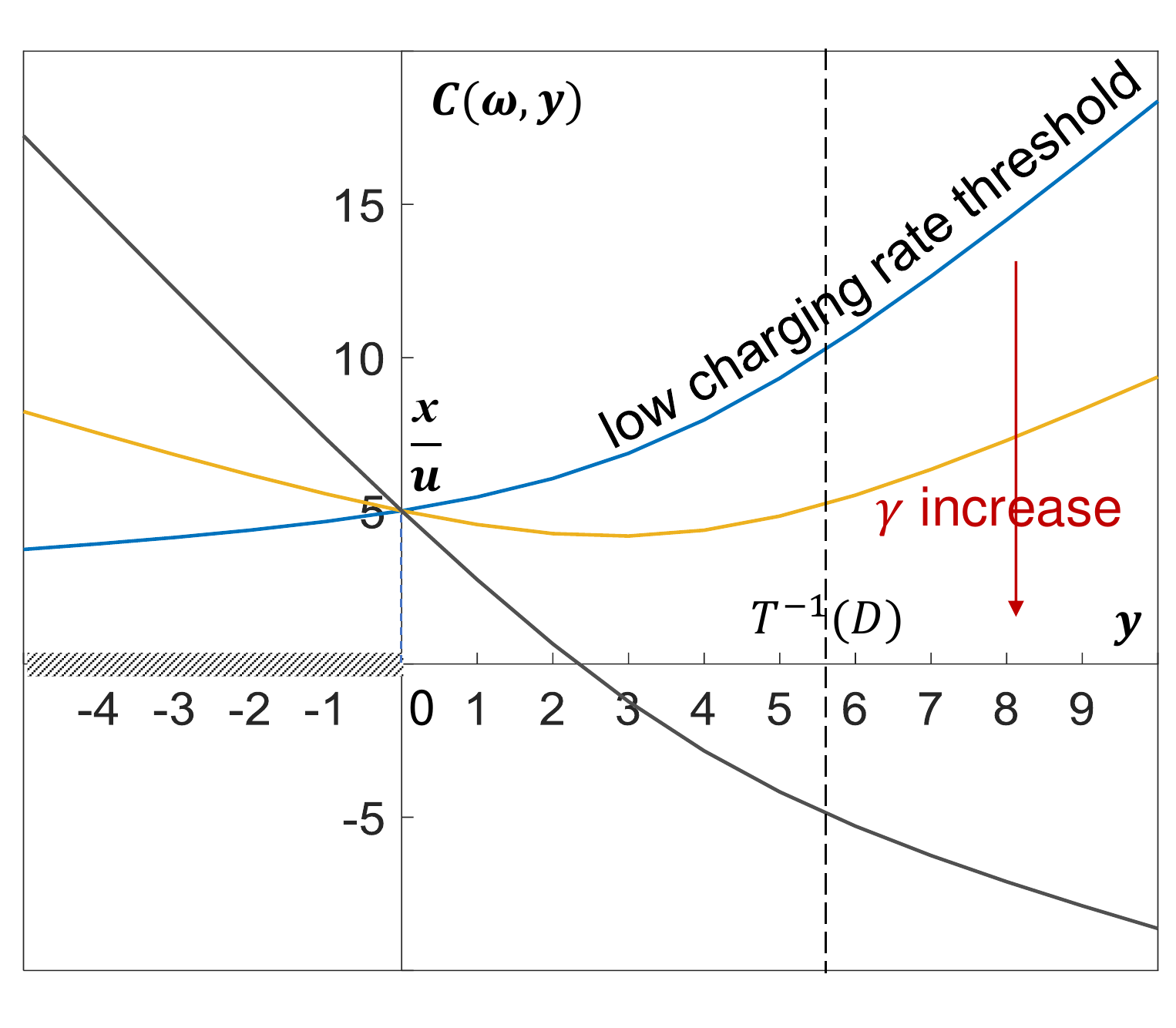}%
\caption{Charging rate $\gamma$ and $C(\omega,y)$. A hitching-plus-charging vehicle with a charging rate lower than the threshold is not eligible. The parameters are $x = 5 \, km, u/v = 2$ and $\omega = 0.3$ with the low charging rate threshold $\gamma = 0.2$.}
\label{fig:chargingrate}
\endminipage\hfil
\minipage{0.32\textwidth}
\centering
\includegraphics[width=0.98\linewidth, height = 1.6in]{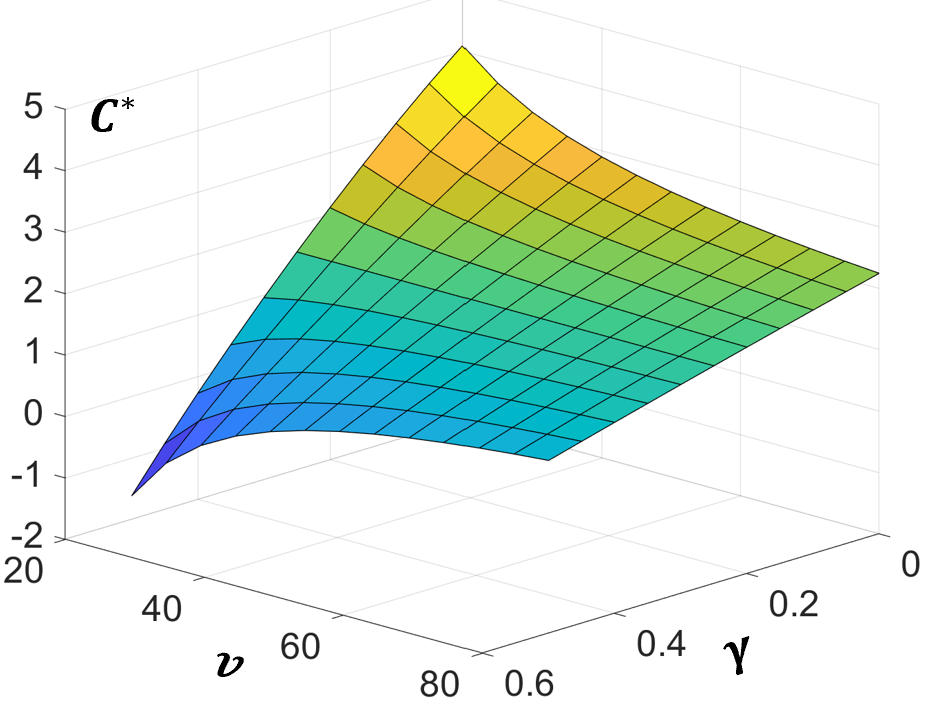}
\vspace{2mm}
\caption{An illustration on $C^*$ under different $v$ and $\gamma$. With charging, the UAV may prefer a slower vehicle. The vehicle speed is from $20$ to $80 km/h$ and the rate $\gamma$ is from $0$ to $0.5$. The parameters are $x = 5 km/h$, $u = 60 km/h$ and $\omega = 0.8$.}
\label{fig:comsume}
\endminipage
\end{figure*}

We optimize $y_{ij}$ between UAV $i$ and vehicle $j$, assuming a fixed pairing relationship. We analyze how UAV $i$ decides between the Choice 1 and Choice 2 in Section II, given the vehicle $j$ is hitching-only or offers charging, respectively.

\subsection{Hitching on a hitching-only vehicle}

Lemma 1 specifies the necessary condition on vehicle speed $v_j$ for hitching-only collaboration.

\vspace{1.5mm}
\noindent{\textbf{Lemma 1.}}
\textit{A hitching-only vehicle $j$ is not eligible for hitching by UAV $i$ if $v_j \leq (1-\omega)u_i$.}
\vspace{1.5mm}

\noindent \textit{Proof.} If $v_j \leq (1-\omega)u_i$, we have $C_{ij}(\omega,y_{ij}) > C_{i0}$. Fig. \ref{fig:speed} illustrates Lemma 1. Hitching on such a slow vehicle yields higher consumption than UAV flying by itself. The parameters in the figures of this section always refer to UAV $i$ and vehicle $j$. We omit the index of $i$ and $j$ for conciseness.

Lemma 1 conveys that in hitching-only, a UAV shall not hitch on a vehicle that is slower than a speed threshold. Next, Proposition 1 clarifies the requirement on $\theta_{ij}$:


\vspace{1.5mm}
\noindent{\textbf{Proposition 1 (Hitching-only vehicle eligibility).}
\textit{A hitching-only vehicle $j$ is eligible for hitching by UAV $i$ if and only if $\displaystyle v_j > (1-\omega)u_i$ and  $\displaystyle \theta_{ij} < \phi_{j}^{ho}(\omega)$, where $\phi_{j}^{ho}(\omega)$ is:}}
\begin{equation}
 \phi_{j}^{ho}(\omega) = \arccos\frac{(1-\omega)u_i}{v_j}.
\end{equation}

\noindent \textit{Proof.} The UAV's consumption $C_{ij}(\omega,y_{ij})$ in (3) is convex in $y_{ij}$. Based on Lemma 1, we derive the optimal $y_{ij}^* = x_i\cos\theta_{ij} - x_i\sin\theta_{ij}\cot\phi_{j}^{ho}(\omega)$. For the hitching to occur, we shall have $y_{ij}^* > 0$, which yields $\theta_{ij} < \phi_{j}^{ho}(\omega)$.
\vspace{1.5mm}

The $\phi_{j}^{ho}(\omega) \leq \pi/2$ indicates a UAV will not choose a hitching-only vehicle that carries it farther away from its destination. We note that $\phi_{j}^{ho}(\omega)$ increases with increasing $\omega$. In particular, when $\omega = 1$, the UAV only cares about its energy and thus $\phi_{j}^{ho}(\omega) = \pi/2$. 

\vspace{1.5mm}
\noindent{\textbf{Proposition 2 (Optimal hitching-only distance).}
\textit{Given a hitching-only vehicle $j$ to UAV $i$, the optimal hitching distance $y_{ij,ho}^{*}$ is:}}
\begin{equation}
y_{ij,ho}^{*}= \begin{cases}
\min\{x_i\cos\theta_{ij}-x_i\sin\theta_{ij}\cot\phi_{j}^{ho}(\omega),T_{ij}^{-1}(D_i)\}, \\
\,\,\,\,\,\quad\quad\quad\quad \text{if $v_j > (1-\omega)v_u$ and $\theta_{ij} \leq \phi_{j}^{ho}(\omega)$}, \\
0, \quad\quad\quad\quad \text{otherwise}.
\end{cases}
\end{equation}

\noindent \textit{Proof.} Given $D_i$, $T_{ij}^{-1}(D_i)$ is the maximum hitching distance allowed in direction $\theta_{ij}$. Following the proof of Proposition 1, the UAV will hitch either to the  optimal location at $x_i\cos\theta_{ij} - x_i\sin\theta_{ij}\cot\phi_{j}^{ho}(\omega)$ or $T_{ij}^{-1}(D_i)$ when it faces $D_i$. 

Since $\phi_{j}^{ho}(\omega)\leq \pi/2$, the $y_{ij,ho}^*$ is no larger than $x_i\cos\theta_{ij}$. When $\omega = 1$, $y_{ij,ho}^* = x_i\cos\theta_{ij}$. This implies the UAV who only cares about energy will hitch to the location closest to its destination. Else, it departs from hitching before reaching $x_i\cos\theta_{ij}$. Substituting $y_{ij} =y_{ij,ho}^*$ into (3), the $C_{ij,ho}^*$ is:
\begin{equation}
C_{ij,ho}^*= \begin{cases}
\frac{x_i}{u_i}\cos(\phi_{j}^{ho}(\omega) -\theta_{ij}),  \text{if $y_{ij} = x_i\cos\theta_{ij} -$}\\
\,\,\,\,\,\quad\quad\quad\quad\quad\quad\quad\quad \text{$x\sin\theta_{ij}\cot\phi_{j}^{ho}(\omega)$}, \\
D_i-\frac{\omega T_{ij}^{-1}(D_i)}{v_j},  \,\,\,\,\,\,\,\,\,\,\,\, \text{if $y_{ij} = T_{ij}^{-1}(D_i)$}, \\
\frac{x_i}{u_i}, \,\,\,\,\,\,\,\quad\quad\quad\quad\quad\quad \text{if $y_{ij} = 0$}. 
\end{cases}
\end{equation}
From (8), a smaller $\theta_{ij}$ reduces the UAV's consumption. With multiple hitching-only vehicles in the same speed, the UAV would prefer the vehicle travelling closest to its destination.

\subsection{Hitching on a hitching-plus-charging vehicle}

We solve UAV $i$'s choice if vehicle $j$ offers charging with a positive rate $\gamma_j$. The results in Subsection III-A are special cases of the results here, with $\gamma_j = 0$. Lemma 2 specifies the necessary condition on $\gamma_j$ for hitching collaboration.

\vspace{1.5mm}
\noindent \textbf{Lemma 2.} \textit{A hitching-plus-charging vehicle $j$ is not eligible for hitching by UAV $i$ if $\omega\gamma_j \leq 1-\omega -\frac{v_j}{u_i} $.}
\vspace{1.5mm}

\noindent \textit{Proof.} When $\omega\gamma_j \leq 1-\omega -\frac{v_j}{u_i} $, the UAV will not benefit from hitching. Fig. \ref{fig:chargingrate} illustrates the threshold of charging rate. A UAV won't hitch on a vehicle with a charging rate lower than the threshold. 

In Proposition 3, we specify the requirement on $\theta_{ij}$:

\vspace{1.5mm}
\noindent{\textbf{Proposition 3 (Hitching-plus-charging vehicle eligibility).}
\textit{A hitching-plus-charging vehicle $j$ is eligible for hitching by UAV $i$ if and only if $\omega\gamma_j \leq 1-\omega -\frac{v_j}{u_i}$ and  $\displaystyle \theta_{ij} < \phi_{j}(\omega,\gamma_j)$, where $\phi_{j}(\omega,\gamma_j)$ is:}}
\begin{equation}
\phi_{j}(\omega,\gamma_j) = \begin{cases}  \arccos\frac{(1-(1+\gamma_j)\omega)u_i}{v_j}, & \text{if $ \omega\gamma_j < 1-\omega +\frac{v_j}{u_i}$},\\
    \pi, & \text{if $ \omega\gamma_j \geq 1-\omega + \frac{v_j}{u_i}$}.
   \end{cases}
\end{equation}

\noindent \textit{Proof.} When $ \omega\gamma_j < 1-\omega +\frac{v_j}{u_i}$, the proof procedure is the same as that in Proposition 1. When $ \omega\gamma_j \geq 1-\omega +\frac{v_j}{u_i}$, hitching can always benefit the UAV regardless of $\theta_{ij}$ due to the fast charging. The bottom curve in Fig. \ref{fig:chargingrate} illustrates this case. We hence derive $\phi_{j}(\omega,\gamma_j) = \pi$ in the second case in (9).

Compared to $\phi_{j}^{ho}(\omega)$ in (6), Proposition 3 presents:
\begin{itemize}
    \item $\phi_{j}(\omega,\gamma_j)$ is increasing in $\gamma_j$. Thus, we have $\phi_j(\omega, \gamma_j) > \phi_j(\omega, 0) = \phi_j^{ho}(\omega)$. 

    \item Unlike $\phi_{j}^{ho}(\omega) \leq \pi/2$, when vehicle $j$ supports charging, $\phi_{j}(\omega,\gamma_j)$ exceeds $\pi/2$ if $1-\omega < \omega\gamma_j$, i.e., when the charging rate is fast enough.
    
    \item \textbf{Hitching on the opposite direction:} When $\gamma_j$ satisfies $ 1-\omega + \frac{v_j}{u_i} \leq \omega\gamma_j$, hitching in any direction $\theta_{ij}$ benefits the UAV, and thus $\phi_{j}(\omega,\gamma_j) = \pi$ in (8). This indicates a UAV may choose a hitching-plus-charging vehicle that travels in the opposite direction to its destination. 
\end{itemize}

Next, we characterize the optimal $y_{ij}^*$ for any value of $\gamma_j$:

\vspace{1.5mm}
\noindent{\textbf{Proposition 4 (Optimal hitching-plus-charging distance).}} \textit{Given a hitching-plus-charging vehicle $j$ to UAV $i$, the optimal hitching distance $y_{ij}^*$ is:} 
\begin{equation}
y_{ij}^{*}= \begin{cases}
\min\{x_i\cos\theta_{ij}-x_i\sin\theta_{ij}\cot\phi_{j}(\omega,\gamma_j),
\\ \quad\quad  T_{ij}^{-1}(D_i)\}, \\
\,\,\, \text{if $\omega\gamma_j > 1-\omega-\frac{v_j}{u_i}$ and $\theta_{ij} \leq \phi_{j}(\omega,\gamma_j)$}, \\
0, \quad \text{otherwise}.
\end{cases}
\end{equation}

\noindent \textit{Proof.} The proof process of Proposition 4 is the same as that in Proposition 2, which we do not repeat.

Compare $y_{ij,ho}^*$ in (6) and $y_{ij}^*$ in (9), we have:

\begin{itemize}
    \item In the same direction $\theta_{ij}$ and speed $v_j$, $y_{ij}^*\geq y_{ij,ho}^*$. The UAV prefers a longer hitching distance on a hitching-plus-charging vehicle. 
    \item \textbf{Hitching over the closest location to the destination:} In hitching-only, the UAV won't hitch over $x_i\cos\theta_{ij}$. If the vehicle $j$  offers charging with rate $\omega\gamma_j > 1-\omega$, the hitch distance $y_{ij}^*$ is longer than $x_i\cos\theta_{ij}$.
\end{itemize}

Using (10), we derive the minimum consumption $C_{ij}^*$:
\begin{equation}
C_{ij}^*= \begin{cases}
\frac{x_i}{u_i}\cos(\phi_{j}(\omega,\gamma_j) -\theta_{ij}), \,\,\,\, \text{if $y_{ij} = x\cos\theta_{ij} -$}\\
\,\,\,\,\,\quad\quad\quad\quad\quad\quad\quad\quad\quad \text{$x\sin\theta_{ij}\cot\phi_{j}(\omega,\gamma_j)$}, \\
D_i-\frac{(1+\gamma_j)\omega T_{ij}^{-1}(D_i)}{v_j}, \,\,\,\,\quad \text{if $y_{ij} = T_{ij}^{-1}(D_i)$}, \\
\frac{x_i}{u_i}, \,\,\,\,\,\,\,\,\,\,\,\,\,\,\quad\quad\quad\quad\quad\quad\quad \text{if $y_{ij} = 0$}.
\end{cases}
\end{equation}

The results in (9)-(11) capture the vehicle features including direction, speed, and charging rate in general.  Fig. \ref{fig:comsume} illustrates the impact of vehicle speed and charging rate on $C_{ij}^*$. Depending on $\gamma_j$, the UAV may prefer to hitch on a slower vehicle for the benefit of charging.

When there are multiple vehicles available, the UAV $i$ can compare $C_{ij}^\ast$ for different vehicle $j$ and determine which one to hitch on. Corollary 1 provides a general rule on UAV $i$'s choice between two heterogeneous vehicles. 

\vspace{1.5mm}
\noindent \textbf{Corollary 1 (Selection between vehicles).} \textit{Consider a UAV $i$ and two vehicles $l, k \in \mathcal{J}$ that are eligible for hitching. When $y_{il}^* < T_{il}^{-1}(D_i)$ and $y_{ik}^* < T_{ik}^{-1}(D_i)$, the UAV $i$ prefers vehicle $k$ if and only if:}
\begin{equation}
    \theta_{ik} - \theta_{il} < \phi_{k}(\omega,\gamma_k) - \phi_{l}(\omega,\gamma_l).
\end{equation}

\begin{figure}
    \centering
    \includegraphics[width=0.9\linewidth]{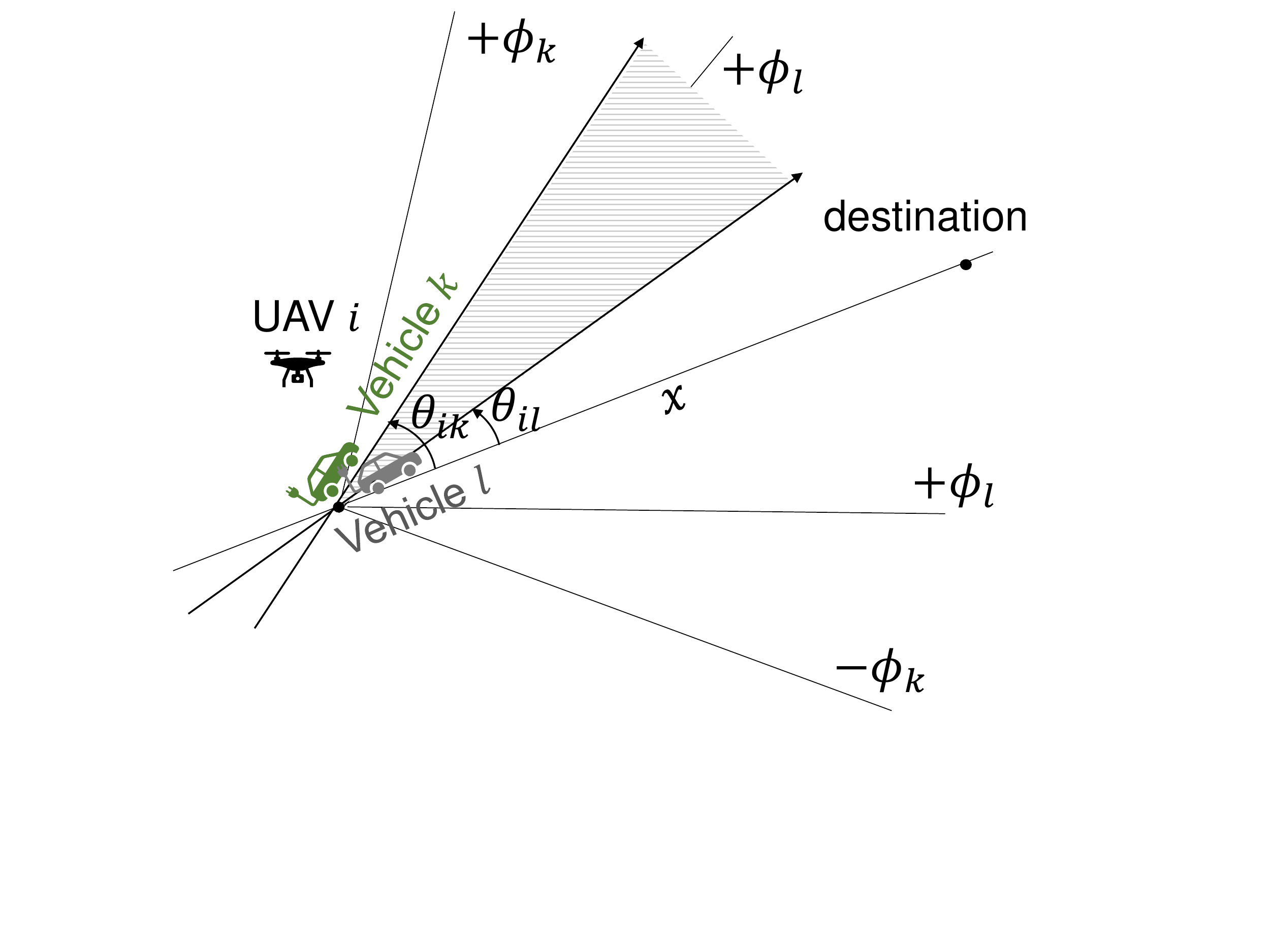}
    \vspace{-45pt}
    \caption{Selection between heterogeneous vehicle $l$ and vehicle $k$. The UAV prefers vehicle $k$ if $\theta_{ik}-\theta_{lk} < \phi_{k}-\phi_{l}$.}
    \label{fig:selection}
\end{figure}

\noindent \textit{Proof.} Using (11), we derive Corollary 1 by comparing $C_{il}^\ast = \frac{x_i}{u_i}\cos(\phi_{l}(\omega,\gamma_l)-\theta_{il})$ and $C_{ik}^\ast = \frac{x_i}{u_i}\cos(\phi_{k}(\omega,\gamma_k)-\theta_{ik})$. When $\theta_{ik} - \theta_{il} < \phi_{k}(\omega,\gamma_k) - \phi_{l}(\omega,\gamma_l)$, we have $C_{il}^* > C_{ik}^*$. The vehicle $k$ is a better choice. Fig. \ref{fig:selection} shows such a selection.

Corollary 1 conveys that the UAV can compare vehicles' direction difference for hitching when $D_i$ is large. Note that when the UAV faces a small $D_i$, the choice will depend on $D_i$. We leave the analysis of $D_i$ in Appendix B.

\section{Stage II: Optimal UAV-Vehicle Collaboration Matching}

Considering $I$ UAVs and $J$ vehicles, solving Problem P1 is equivalent to solve the following Problem P2:
\begin{subequations}
\begin{alignat}{4}
\text{P2:} \,\, & \underset{\bm{b}}{\text{maximize}}   &\,\,& \sum_{i=1}^{I}\sum_{j=1}^{J} (C_{i0} - C^*_{ij}) b_{ij}\\
& \text{subject to}   &\,\,& \sum_{i=1}^{I} b_{ij} \leq 1, \forall j; \sum_{i=1}^{J} b_{ij} \leq 1, \forall i,\\
& &\,\,&  b_{ij}\in\{0,1\},\forall i,j,
\end{alignat}
\end{subequations}
where $C^*_{ij}$ is characterized in (11), and $C_{i0} -C^*_{ij}$ corresponds to how much (time and energy) consumption saving by UAV $i$ via hitching on vehicle $j$. Therefore, in Problem P2, the only variables optimized are $\{b_{ij}\}$, indicating which UAV should hitch on which vehicle. The constraints are inherited from the original Problem P1.

Problem P2 is a maximum weight bipartite graph matching (BGM) problem. For BGM, a maximum weight matching is the one that maximizes the sum of the weights of the edges between matched vertex nodes \cite{BM}. Finding an optimal matching follows the primal-dual framework. We develop the Max-Saving Algorithm (MSA) to solve Problem P2.

\begin{algorithm}
\DontPrintSemicolon
\caption{Max-Saving UAV-vehicle matching}


\KwInput{UAV and vehicle nodes, $U$ and $V$, UAVs' $\{(x_i, u_i, D_i)\}$, vehicles' $\{(\theta_{ij}, v_j,\gamma_j)\}$}

Compute $\{(C_{i0} - C_{ij}^*)\}$ using (11)

G $\leftarrow \emptyset$ \quad\quad \tcp{optimal matching} 

$p_i \leftarrow \max_{j\in V} (C_{i0} - C_{ij}^*)$ for UAV $i$;

$q_i \leftarrow 0$  for vehicle $j$;

\While{U $\neq \emptyset$ or $\exists i, p_i\neq 0$}
{
  Adjust G based on edges in $p_i + q_j =  (C_{i0} - C_{ij}^*)$\\
  Add reachable UAV-vehicle nodes to $U'$ and $V'$\\
  \For{vehicle $m \in V \setminus V'$ and UAV $n \in U'$ }
  {$\varepsilon = \min\{p_n + q_m -  (C_{n0} - C_{nm}^*), 0\}$}
  
  \For{vehicle $m \in V'$ and UAV $n \in U'$}
  {  $p_n := p_n - \varepsilon $; $q_m := q_m + \varepsilon $}
}
\textbf{return:} G and the total savings $W(\text{G})$.
\end{algorithm}

Algorithm 1 summarizes the MSA. Based on the parameters $\{(x_i,u_i,D_i)\}$ of UAVs and $\{(\theta_{ij}, v_j,\gamma_j)\}$ of vehicles, we compute $\{C_{i0} - C_{ij}^*\}$ using (11), which serve as the edge weights (Line 1). We initialize a graph $G$ to include matched nodes (Line 2). The variables $p_i$ and $q_j$ act as the primal and dual pairs for updating the matching (Lines 3 to 4). In each iteration, we add new UAV-vehicle pairs to $G$ based on $p_i$ and $q_j$ (Lines 6 to 7). Then, variables are updated (Lines 8 to 13). The iteration terminates when all UAVs are accommodated or no unmatched UAV can be added to $G$. We conclude the optimality of Algorithm 1 in Theorem 1:

\vspace{1.5mm}
\noindent \textbf{Theorem 1:} \textit{The Max-Saving Algorithm computes the global optimal solution of Problem P2 with a time complexity of  $\mathcal{O}(I^2J)$.}
\vspace{1.5mm}

\noindent \textit{Proof.} The time complexity depends on the number of UAVs, $I$, times the number of all possible UAV-vehicle pairs, $IJ$. We provide the proof details in the appendix \cite{TR}.

\section{Performance Analysis}
\begin{figure}[!t]
\subfloat[Case 1: $\theta_{ij}\leq\pi$]{\includegraphics[width=0.48\linewidth]{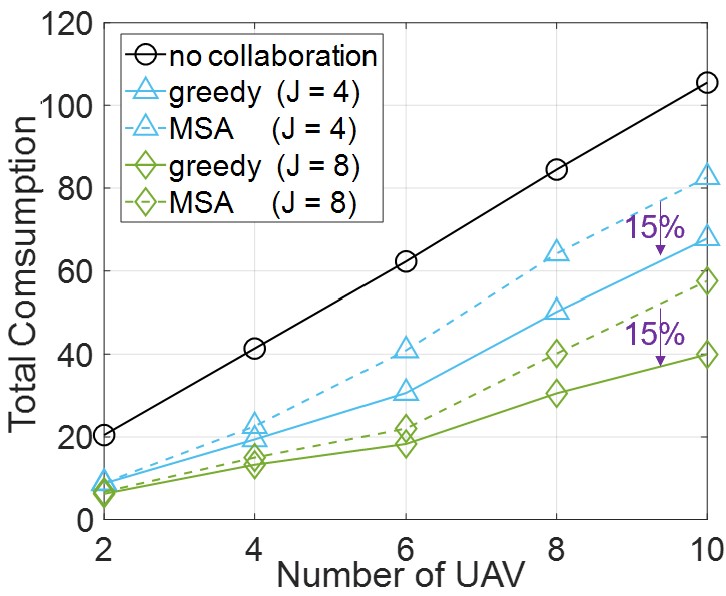}%
\label{fig:case1}}
\hfil
\subfloat[Case 2: $\theta_{ij}\leq\pi/2$]{\includegraphics[width=0.48\linewidth]{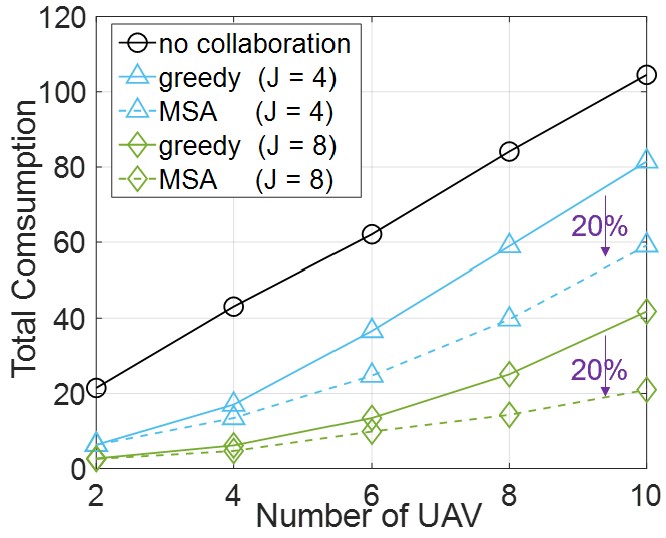}%
\label{fig:case2}}
\caption{Comparison of UAVs' total consumption using the MSA and the greedy matching.}
\label{fig:SMA}
\vspace{-5pt}
\subfloat[Case 1: $\theta_{ij}\leq\pi$]{\includegraphics[width=0.49\linewidth]{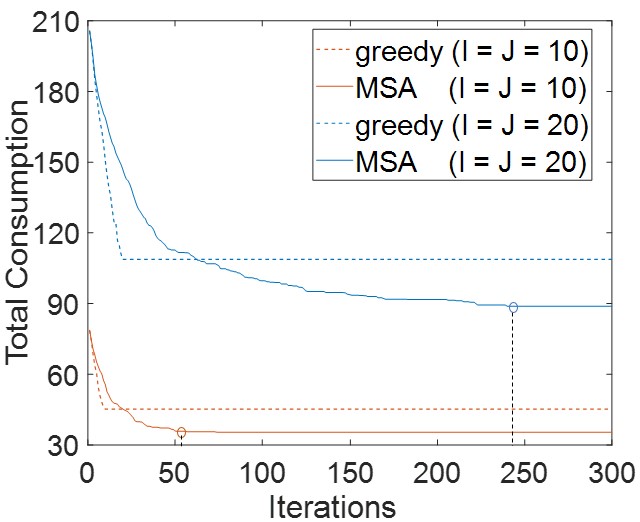}%
\label{fig:case11}}
\hfil
\subfloat[Case 2:  $\theta_{ij}\leq\pi/2$]{\includegraphics[width=0.49\linewidth]{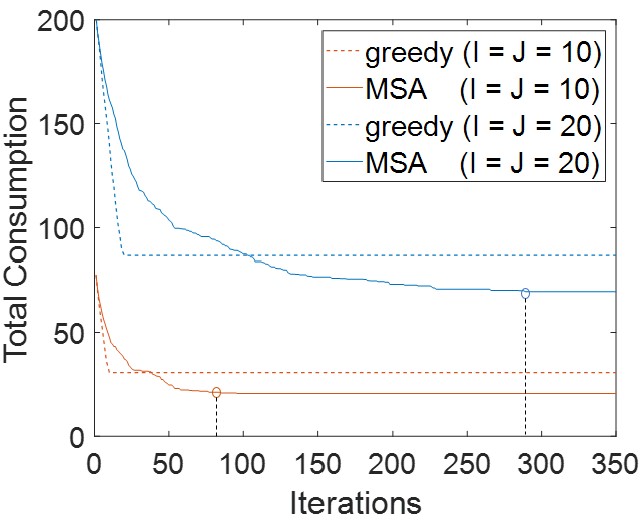}%
\label{fig:case21}}
\caption{Performance of the MSA convergence and optimality. Simulations are repeated 100 times}
\label{fig:Converge}
\end{figure}
\setlength{\textfloatsep}{10pt}

We perform extensive simulations to evaluate the optimal UAV hitching strategies considering different numbers of UAVs and vehicles. The UAV distance $x_i$ is random within $20 \, km$. We set $\omega= 0.8$ and $u_i = 60 \, km/h$. The vehicles have speed $v_j = 40 \, km/h$ and charging rate $\gamma_j = 0.3$. We simulate: (1) Case 1: the direction $ \theta_{ij}$ is random within $[0, \pi]$; (2) Case 2: the direction $ \theta_{ij} $ is random within $[0, \pi/2]$.

Fig. \ref{fig:SMA} demonstrates the effectiveness of using MSA in matching UAV-vehicle collaboration over using the heuristic greedy matching. We compare the consumption of UAVs by these two methods with the benchmark, where UAVs fly to their destinations. Fig. \ref{fig:SMA} shows without hitching, the total consumption of UAVs linearly increases with the UAV population. In comparison, the MSA effectively reduces the consumption under different $\theta_{ij}$ ranges and vehicle numbers as shown in Fig. \ref{fig:case1} and Fig. \ref{fig:case2}. Compared with the greedy method, the MSA achieves more saving for UAVs due to the optimal UAV-vehicle collaboration matching. When the number of UAVs increases, results show $15\%$ and $20\%$ more consumption reduction by the MSA against the greedy method in Case 1 and Case 2, respectively. 

Fig. \ref{fig:Converge} presents the number of iterations executed for the MSA to complete an optimal matching. Compared to the greedy method, the advantage of the MSA is that it minimizes the UAVs' consumption. The iteration times in our simulations are substantially less than the theoretical value $I^2J$. Part of the reason is that each UAV only needs to consider the vehicles eligible for it to hitch on. This eliminates ineligible choices and hence reduces the time needed to reach an optimal matching. Note that as the range of $\theta_{ij}$ is larger in Case 1, there would be fewer eligible vehicles for a UAV to consider than in Case 2. Thus, comparing Fig. \ref{fig:case11} and Fig. \ref{fig:case21}, the number of iterations increases in Case 2. The benefit is that as more vehicles are eligible for hitching in Case 2, more saving is achieved for UAVs as Fig. 6 shows.

\section{Conclusions}
This paper presented the first theoretical study on UAVs' optimal hitching strategies on ground vehicles to save their energy. Given ground vehicles' travelling pattern and charging capabilities, we reported that depending on the charging rate, the UAV may hitch on a vehicle that carries it farther away from its destination. The UAV may also choose a slower vehicle to hitch, and prolong the hitching distance for the benefit of battery charging. We derived that among heterogeneous vehicles, the UAV can make a choice based on the vehicle direction difference. For multiple UAVs, extensive simulations demonstrated that our MSA optimizes the UAV-vehicle collaboration and maximizes the UAVs' benefits. Note that we studied UAVs' hitching given a particular traffic state of the vehicle in this paper. We plan to study the hitching collaboration under dynamics of ground traffic conditions.

\appendices
\section{Optimal UAV Hitching (Limited Battery Capacity)}

We consider the UAV's battery capacity is $e_{full}$. The energy state of UAV $i$ when it starts hitching is denoted by $e_i$. Therefore, during hitching, the UAV $i$ at most can be recharged with $\Delta e_i = e_{full}-e_i$ amount of energy.  First, we note that UAV's energy state does not affect the conditions of vehicle eligibility specified in Proposition 1 and Proposition 3.  

Considering $e_{full}$ and $e_i$, the energy consumption of UAV $i$ in hitching a distance $y_{ij}$ on vehicle $j$ is:
\begin{equation}
\begin{aligned}
E_{ij,lim}(y_{ij}) = 
\displaystyle  \frac{\sqrt{x_i^2-2x_iy_{ij}\cos\theta_{ij}+y_{ij}^2}}{u_i} \\
-\min\{e_{full} - e_i, \frac{\gamma_j}{v_j}y_{ij}\}.
\end{aligned}
\end{equation}

On the right side of (14), the term, $\min\{e_{full} - e_i, \frac{\gamma_j}{v_j}y_{ij}\}$, indicates during hitching, the energy charged to the UAV cannot exceed $e_{full} - e_i$. The consumption of UAV is expressed as:
\begin{equation}
C_{ij,lim}(\omega,y_{ij}) = \omega E_{ij,lim}(y_{ij}) + (1-\omega)T_{ij}(y_{ij}).
\end{equation}

By optimizing the $y_{ij}$ to minimize (15), we derive UAV $i$'s optimal hitching distance, $y_{ij,lim}^*$, on an eligible vehicle $j$, which is specified in Proposition 5.

\vspace{2mm}
\noindent{\textbf{Proposition 5 (Optimal hitching distance under limited UAV battery capacity).}} \textit{Given an eligible vehicle $j$ to UAV $i$, the optimal riding distance $y_{ij,lim}^*$ considering UAV's battery capacity capacity $e_{full}$ and energy state $e_i$ is as follows:} 
\begin{equation}
y_{ij,lim}^*= \begin{cases}
\displaystyle y_{ij}^*, \quad\quad\quad \text{if $\frac{(e_{full} - e_i)v_j}{\gamma_j} \geq  y_{ij}^*$},\\
\displaystyle y_{ij,ho}^*, \quad\quad \text{if $\frac{(e_{full} - e_i)v_j}{\gamma_j} \leq  y_{ij,ho}^*$}, \\
\displaystyle \min\{\frac{(e_{full} - e_i)v_j}{\gamma_j} , T_{ij}^{-1}(D_i)\}, \text{otherwise}.
\end{cases}
\end{equation}

\begin{figure}[t]
\centering
\includegraphics[width=0.6\linewidth]{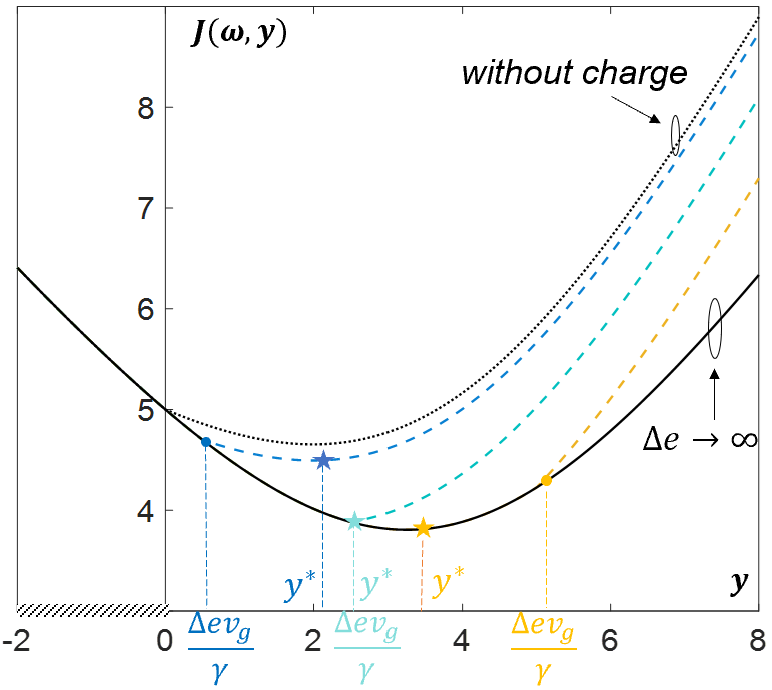}%
\caption{$y^*$ on hitching-plus-charging vehicle under different $\Delta e$ and $D$.}
\footnotesize{($x = 5 \, km, u = 60 \, km/h, v = 30 \, km/h, \theta =\pi/4$)}
\label{fig:bc}
\end{figure}

\noindent \textit{Proof.} We note that $y_{ij,ho}^{*}$ is UAV $i$'s optimal hitching-only distance, expressed in (7), and $y_{ij}^*$ is the optimal hitching-plus-charging distance under the assumption of infinite UAV battery capacity, expressed in (10). Fig. \ref{fig:bc} illustrates the $y_{ij,lim}^*$ in (16).

First, if the UAV cannot be fully charged when it hitches to $y_{ij}^*$, i.e., $\frac{(e_{full} - e_i)v_j}{\gamma_j} \geq  y_{ij}^*$, the situation is equivalent to when we assume UAV $i$ has large battery capacity and can receive as much as energy as it wants during hitching. Thus, the optimal hitching distance is exactly $y_{ij}^*$ as shown in Fig. \ref{fig:bc}. This corresponds to the upper most case in (16).

Second, if the UAV is fully recharged before reaching $y_{ij,ho}^*$, i.e., $\frac{(e_{full} - e_i)v_j}{\gamma_j} \leq  y_{ij,ho}^*$, the UAV shall continue hitching to $y_{ij,ho}^*$. This is because once the battery is fully recharged, the hitching that continues is as the UAV is hitching on a hitching-only vehicle. Hence, the $y_{ij,ho}^*$ is the optimal distance as shown in Fig. \ref{fig:bc}. This corresponds to the middle case in (16).

Finally, if the UAV is fully charged at any distance between $y_{ij,ho}^*$ and $y_{ij}^*$, i.e.,  $y_{ij,ho}^* < (e_{full} - e_i)v_j/\gamma_j <  y_{ij}^*$, the UAV shall depart from hitching at the time the battery is full. This is because the UAV will no longer benefit from charging once it is fully charged. Fig. \ref{fig:bc} illustrates this case.
\vspace{2mm}

From Proposition 5, we analyze UAV's hitching decision when $\gamma_j \rightarrow \infty$, i.e., the UAV can swap a new battery on the vehicle. When $\gamma_j \rightarrow \infty$, we have $(e_{full} - e_i)v_j/\gamma_j \rightarrow 0$ and eligible direction range $\phi_{ij}(\omega,\infty) = \pi$ based on Proposition 3. As such, we derive \textbf{UAV's hitching strategy for battery swapping:}
\begin{itemize} \itemsep 3pt
    \item Any direction $\theta_{ij} \leq \pi$ is an eligible direction. The UAV can hitch on any vehicle that is available, regardless of the vehicle speed and travel direction.
    \item Further, if $\theta_{ij} \leq \phi_{ij}^{ho}(\omega)$, the optimal hitching distance is $y_{ij,ho}^*$. This indicates after battery swapping, the UAV shall continue hitching to the optimal hitching-only distance.
    \item If $\theta_{ij} > \phi_{ij}^{ho}(\omega)$, the vehicle can offer a new battery, but is ineligible for hitching-only. Thus, the UAV shall immediately depart from the vehicle once it gets a new battery.
\end{itemize}

\section{Heterogeneous Vehicle Selection under Different Time Restrictions}

In Corollary 1, we specify that when the UAV has a large $D_i$, it can simply compare vehicles based on the different between vehicles' directions.  In the following Corollary, we explain the UAV's choice when it can only hitch on vehicles for limited amount of time:

\vspace{2mm}
\noindent \textbf{Corollary 2 (Selection between vehicles under time restriction).} \textit{Consider a UAV $i$ and two vehicles $l, k \in \mathcal{J}$ that are eligible for hitching. When $y_{il}^* = T_{il}^{-1}(D_i)$ and $y_{ik}^* = T_{ik}^{-1}(D_i)$, the UAV $i$ prefers vehicle $k$ if and only if:}
\begin{equation}
\displaystyle   T_{ik}^{-1}(D_i) > \frac{\{1-(1+\gamma_l)\omega\}v_k}{\{1-(1+\gamma_k)\omega\}v_l}T_{il}^{-1}(D_i).
\end{equation}

\noindent \textit{Proof.} if UAV $i$ hitches on vehicle $l$, the UAV's minimum consumption is $C_{il}^* = D_i - \frac{1-(1+\gamma_l)\omega}{v_l}T_{il}^{-1}(D_i)$. Likewise, if hitching on vehicle $k$, the consumption is $C_{ik}^* = D_i - \frac{1-(1+\gamma_k)\omega}{v_k}T_{ik}^{-1}(D_i)$. Setting $C_{ik}^* < C_{il}^*$ yields the condition in Corollary 2.

Recall that give a vehicle $j$ to UAV $i$, $T_{ij}^{-1}(D_i)$ is the maximum allowed hitching distance under the time restriction $D_i$. We derive $T_{ij}^{-1}(D_i)$ as follows:
\begin{equation}
\begin{aligned}
&T_{ij}^{-1}(D_i) = (1-\frac{u_i^2}{v_j^2})(x_i\cos\theta_{ij} - \frac{D_iu_i^2}{v_j})+\\
&(1-\frac{u_i^2}{v_j^2})\sqrt{(\frac{u_i^2}{v_j^2}-1)x^2\sin^2\theta_{ij} + (\frac{u_i}{v_j}x_i\cos\theta_{ij} - D_iu_i)^2}
\end{aligned}
\end{equation}

From Corollary 2, if the UAV faces an urgent task, it needs to make a choice based on the $D_i$. The UAV inclines to choose the vehicle travelling in a closer direction to its destination when $D_i$ reduces.

\section{Extension of the MSA for Vehicle Capacity Heterogeneity}

\subsection{Proof of Theorem 1}

\vspace{2mm}
\noindent \textbf{Theorem 1:} The Max-Saving Algorithm computes the global optimal solution of Problem P2 with a time complexity of  $\mathcal{O}(I^2J)$. 
\vspace{2mm}

\noindent \textit{Proof.} \textbf{Optimality:} The MSA updates the matching based the primal-dual framework. The $p_i$ and $q_j$ represent the primal and dual variable associated with UAV $i$ and vehicle $j$, respectively. The MSA optimizes the dual problem of Problem P2:
\begin{equation}
\begin{aligned}
\underset{\bm{p},\bm{q}}{\text{maximize}}  \sum_{i=1}^{I}\sum_{j=1}^{J} (C_{i0} - C^*_{ij}) b_{ij} - \sum_{i=1}^{I}\sum_{j=1}^{J} p_ib_{ij} - \sum_{i=1}^{I}\sum_{j=1}^{J} q_jb_{ij}.
\end{aligned}
\end{equation}

Note that Problem P2 is an integer linear programming. Optimally solving the dual problem optimizes the original problem. Therefore, the MSA globally maximizes the benefits for UAVs. The optimality of the MSA is proved. We note that depending on the parameters of UAVs and vehicles, there could exists multiple optimal matchings that leads to the same unique global maximum benefits. In this case, the MSA always finds an optimal matching within these matchings.

\textbf{Time complexity:} For the time complexity, each iteration of Algorithm 1 adds at least one node to the matching. Thus, the iteration occurs $\mathcal{O}(I)$ times. Within each iteration, the algorithm needs at most  $\mathcal{O}(IJ)$ times of primal-dual variables updating. Then, it takes $\mathcal{O}(IJ)$ time to
adjust the matching by finding an alternating path, which increases the size and weights sum of the matching \cite{TR}. Specifically, in our context, an alternating path starts from an unmatched nodes and traverses the path through tight edges, i.e., edge weights satisfy $p_i + q_i = C_{i0} - C_{ij}^*$ in and outside the matched graph alternatively. In this way, adding an unmatched node and adjusting the matching are completed simultaneously. Since there are $IJ$ entries, finding such a path has a time complexity $\mathcal{O}(IJ)$.  Thus, the time needed within each iteration is $\mathcal{O}(IJ)$. Overall, the total time complexity of the MSA is $\mathcal{O}(I^2J)$.

\subsection{Heterogeneous Vehicle Capacity}
Consider a vehicle $j\in \mathcal{J}$ can carry $z_j$ number of UAVs. The optimal UAV-vehicle collaboration is formulated as:
\begin{subequations}
\begin{alignat}{4}
\text{P2:} \,\, & \underset{\bm{b}}{\text{maximize}}   &\,\,& \sum_{i=1}^{I}\sum_{j=1}^{J} (C_{i0} - C^*_{ij}) b_{ij},\\
& \text{subject to}   &\,\,& \sum_{i=1}^{I} b_{ij} \leq z_j, \forall j,\\
& &\,\,& \sum_{i=1}^{J} b_{ij} \leq 1, \forall i,\\
& &\,\,&  b_{ij}\in\{0,1\},\forall i,j,
\end{alignat}
\end{subequations}
where constraint (20b) ensures that vehicle $j$ accommodates $z_j$ number of UAVs at most. The objective and the rest constraints remain unchanged as in Problem P2. Note that considering vehicle capacity $z_j$ does not change the structure of the original problem. We present the extended MSA in Algorithm 2, which optimally solves Problem P3.

Compared to Algorithm 1, the additional operation in the extended MSA is that in Line 1, we create $z_j - 1$ number of virtual nodes for each vehicle $j$. The connections of these virtual nodes to UAVs are identical to how the vehicle $j$ is connected to the UAVs. After this initialization, the following procedure is the same as the Algorithm 1. Based on Theorem 1, we can derive the optimality and time complexity of the extended MSA as follows:

\vspace{2mm}
\noindent \textbf{Corollary 3.} \textit{The Extended Max-Saving Algorithm computes the global optimal solution of Problem P3 with a time complexity of  $\mathcal{O}(I^2\sum_{j=1}^{J}z_j)$. }
\vspace{2mm}

\noindent \textit{Proof.} Since there are $z_j$ positions on each vehicle $j$, we have $\sum_{j=1}^{J}z_j$ number of vehicle nodes in total in $V$ for matching as Algorithm 2 shows. Thus, the possible number of edges between UAVs and vehicles are $I\sum_{j=1}^{J}z_j$. Based on Theorem 1, the time complexity of the extended MSA is $\mathcal{O}(I^2\sum_{j=1}^{J}z_j)$.

\vspace{1mm}
\begin{algorithm}[h]
\DontPrintSemicolon
\caption{Extended MSA}


\KwInput{UAV and vehicle nodes, $U$ and $V$, UAVs' $\{(x_i, u_i, D_i)\}$, vehicles' $\{(\theta_{ij}, v_j,\gamma_j, z_j)\}$}

Create $z_j-1$ virtual vehicle nodes accompanying vehicle $j$ to $V$

Compute $\{(C_{i0} - C_{ij}^*)\}$ using (11)

G $\leftarrow \emptyset$ \quad\quad \tcp{optimal matching} 

$p_i \leftarrow \max_{j\in V} (C_{i0} - C_{ij}^*)$ for UAV $i$;

$q_i \leftarrow 0$  for vehicle $j$;

\While{U $\neq \emptyset$ or $\exists i, p_i\neq 0$}
{
  Adjust G based on edges in $p_i + q_j =  (C_{i0} - C_{ij}^*)$\\
  Add reachable UAV-vehicle nodes to $U'$ and $V'$\\
  \For{vehicle $m \in V \setminus V'$ and UAV $n \in U'$ }
  {$\varepsilon = \min\{p_n + q_m -  (C_{n0} - C_{nm}^*)\}, 0\}$}
  
  \For{vehicle $m \in V'$ and UAV $n \in U'$}
  {  $p_n := p_n - \varepsilon $; $q_m := q_m + \varepsilon $}
}
\textbf{return:} G and the total savings $W(\text{G})$.
\end{algorithm}


\end{document}